\documentclass[amsmath,amsfonts]{article}
\usepackage{tikz}
\usetikzlibrary{arrows,shapes,trees}
\usepackage{graphicx}
\usepackage{color,soul}
\definecolor{lightblue}{rgb}{.90,.99,1}
\usepackage{color}
\usepackage{float}
\sethlcolor{lightblue}
\def\be{\begin{equation}}
\def\eea{\end{eqnarray}}
\date{}
\usepackage{cite}
\def\bea{\begin{eqnarray}}
\def\ee{\end{equation}}

\author{H.Pahlavani \footnote{ h-pahlavani@qom.ac.ir}, M.Botshekananfard \footnote{mbotshekanan@yahoo.com}
\\ {\small Department of Physics, University of Qom.}}
\title{Two-dimensional electron gas under the effect  of constrained potential and magnetic field in curved space}
\begin{document}
\maketitle
\begin{abstract}
 The effect of the curvature of a cylindrical surface on the energy spectrum for a curved two dimensional electron gas in a homogeneous magnetic field is considered. The corrections to the energy spectrumnis obtained for the first time perturbatively, in contrast to previous works where it was obtained numerically. The dispersion relationship is obtained as  a function of curvature radius and the results for curved surface have been compared with the flat surface.\\\\
 {\bf Keywords: Two-dimensional electron gas, Curvature radius, Constrained potential, Geometric potential} \\\\
 {\bf PACS numbers: 73,43,Fj, 73,23,Ad, 73,43,Qt}
\end{abstract}
\section{Introduction}
In quantum mechanics, one of the ways that can be discussed about
the motion of a particle rigidly bounded on a surface is
confining potential approach, in that the particle is confined
by a strong force that acts normally to our surface in all points
of the space.This idea can be readily put in practice considering
a potential which is constant over the surface but increases
sharply for every small displacement in the normal direction to
the surface. The confining potential approach yields a unique
effective Hamiltonian that depends on physical mechanism of the
constraint  \cite{net1,net2,net3}. The Hamiltonian contains the
surface potential as the quantum potential that is dependent on
mean and Gaussian curvatures\cite{net1,net4,net5}. One of the
greatest  achievements in solid state physics is the fabrication
of low dimensional systems. An example of a low dimensional system
is the two-dimensional electron gas (2DEG). The electrical
behavior of a (2DEG) subjected to a uniform magnetic field has
been studied in much detail \cite{net6,net7,net8}. In the
ballistic regime, the system is characterized by stationary Landau
states. A proemial label  of two dimensional implied that these
electron systems  were flat. The physics of nanostructures and
quantum waveguide may pose questions concerning curved surface in
quantum theory, which is increasingly relevant to device modelling
\cite{net9, net10}. Two-dimensional electron gas (2DEG) in the
planar heterostructures has been investigated greatly, which lead
to in finding  a number of remarkable quantum phenomena such as,
integer and fractional quantum Hall effects, the Berry quantum
phase,etc.  Non-planar
low-dimensional structures are fundamentally new physical objects
that have  attracted  the attention of
the researchers  during last recent years \cite{net8,net11,net12,net13,net14,net10,net15,net16,net17,net18,net19,net20}.\\
 Experimentally non-planar surfaces with 2DEG were
synthesized by means of molecular-beam epitaxy on faceted surfaces
\cite{net21,net22} and were typically realized at nearly
atomically smooth interfaces of single-crystalline semiconductors
\cite{net23}. The main drawback of such structures is spatial
fluctuations of the curvature of surface in them, which makes the
studies of the effect of the curvature on the magnetotransport
\cite{net24} in such structures a very difficult task. The
cylindrical surface is the simplest model for investigating the
influence of geometry on physical systems\cite{net25,net26}. The
interest in the electronic properties of quantum systems with
cylindrical symmetry has received a boost, because the early
proposals of carbon nanotubes \cite{net27, net28} for building
 future nanoelectronic devices, have interesting mechanical and electrical properties.
In this paper, curved two-dimensional electron gas with
cylindrical symmetry was considered  in a homogeneous magnetic
field. The energy spectrum of the electrons on a cylinder surface have been obtained by studying  the geometry potential as the
constraint.
\section{Hamiltonian of a 2DEG in  a magnetic field}
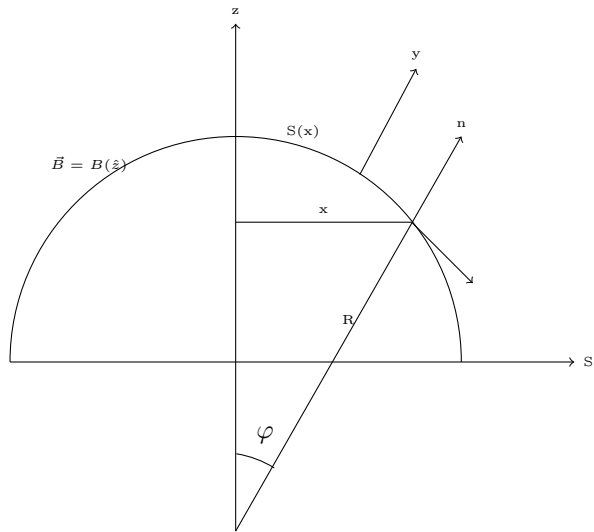
\begin{figure}[H]\label{fig1}
\centering
\begin{tikzpicture}[scale=3]
\draw[->] (-1,0) -- (1.5,0)node[right] {\tiny S};
\draw[->] (0,-.75) -- (0,1.5)node [above]{\tiny z};
\draw[->] (0,-.75) -- (1,1)node [above]{\tiny n}node[pos=0.5,above]{\tiny R};
\draw[->] (.55,.831) -- (.8,1.3)node [above]{\tiny y};
\draw[->] (.78,.62) -- (1.05,.35);
\draw(0,.62) -- (.78,.62)node[pos=0.5,above]{\tiny x};
\draw (0:1) arc (0:180:1);
\fill (-.65,.8) node[above] {\tiny $\vec{B}=B (\hat{z})$};
\fill (.3,.95) node[above] {\tiny S(x)};
\draw (-70:.5) arc (58:81:.45);
\fill (.13,-.4) node[above] {$\varphi$};
\end{tikzpicture}
\caption{ Curved electron gas coordinate system in a homogeneous magnetic field} 
\end{figure}

Let us consider  a spinless electron in two dimensions 
constrained to move along a surface $s$ of curve $C$ (curved
two-dimensional electron gas (C2DEG)) by the action of external
forces. For this purpose, we consider a curved two-dimensional
electron gas (2DEG) in a uniform magnetic field $\vec{B}=B 
\cos(\varphi) \hat{z}$ that is shown in Fig.\ref{fig1}. Without any
restriction on the problem and just for simplicity, we assume that
the equation of the surface is $z=f(x)$, where $f(x)$ is an
arbitrary function which depend on the variable $y$. According to
Fig.\ref{fig1}, the arc length of $C$ (the distance parameter on
the surface).i.e $s$ is defined by
\begin{equation}\label{eq1}
 s(x)=\int_{0}^{x}\sqrt{{1+f^{'}}^{2}(x)}dx,
\end{equation}
 where the coordinate $x(s)$ is function of the arc length of $C$.
 The problem then reduces to solving the Schrodinger equation
 \begin{equation}
 \frac{1}{2m}\Big[({\mathbf{P}}-\frac{e}{c}{\mathbf{A}})^2+V_{\lambda}(z)\Big]\psi(s,y,z)=E\psi(s,y,z).
\label{eq2}
\end{equation}

Since in quantum mechanics, we can no longer predict the position
of the particle with point like accuracy, it is natural to
consider only constraint forces that are orthogonal to the curve $C$
in all points of the plane, where the particle can possibly be
found. In order to satisfy this requirement, we shall consider
potentials which have constant values over $C$ but are increased
sharply for every small displacement along normals of $C$. It can
be easily see that this result may be obtained by choosing
potentials $V_{\lambda}(z)$ (independent of $s$). For this purpose
 the constraining process can be defined as being produced by a family
of increasingly stranger potentials $V_{\lambda}(z)$, where
$\lambda (z)$ is a squeezing parameter which defines the
strength of the potential \cite{net1}
\begin{equation}
 V_{\lambda}(z)=\left\{
\begin{array}{cc}
0   &   z=0\\
\infty   &  z\neq 0
\end{array}\right\}.
\label{eq3}
\end{equation}
Taking the magnetic field $\vec{B}=(0,0,B\cos\phi)$ according  to the
Fig.\ref{fig1}, the vector potential is $\vec{A}=(0,Bx(s),0)$.\\

Eq.(\ref{eq2}) can now be easily separated by setting
$\psi(s,y,z)=Y(s,y).Z(z)$ where $Y(s,y)$ is the tangential component of the wave function
\begin{equation}
\frac{-\hbar ^2}{2m}\frac{\partial ^{2} Z(z)}{\partial{z}^2}+V_{\lambda}(z) Z(z)=i\hbar \frac{\partial  Z(z)}{\partial{t}},
\label{eq4}
\end{equation}
\begin{eqnarray}
&&\frac{-\hbar^2}{2m}\frac{\partial^2{Y(s,y)}}{\partial{s}^2}+\frac{-\hbar^2}{2m}\frac{\partial^2{Y(s,y)}}{\partial{y}^2}+\frac{1}{2}m\omega_{c}^2 x^2(s)Y(y,s)+i\hbar \omega_{c} x(s)\nonumber\\
&&\frac{\partial{Y(y,s)}}{\partial{y}}+U(s,y) Y(y,s)
= E Y(y,s),
\label{eq5}
\end{eqnarray}
where $\omega_{c}=e B/m c$. \\
Eq.(\ref{eq4})  that is just
a one-dimensional Schrodinger equation for a particle bounded by
the transverse potential $ V_{\lambda}(z)$, can be ignored in all
future calculation. Eq. (\ref{eq5}) is much more interesting,
due to the presence of the surface potential $U(s,y)$. In the case
of a 2DEG flexing the gas leads to a geometric potential of the
form \cite{net1}
 \begin{equation}
U=\frac{-\hbar^2}{8m}[\kappa_{1}(s)-\kappa_{2}(s)]^2=\frac{-\hbar^2}{8m}\Big(\frac{1}{R_{1}(s)}-\frac{1}{R_{2}(s)}\Big)^2,
\label{eq6}
\end{equation}
where $m$ is the effective mass  and $R_{1}$,$R_{2}$ are the
principal curvature radii of the surface at the point where the
electron resides\cite{net2, net3}.\\
The surface potential (geometric potential) is always attractive and is independent of the electric charge of particle, similar to gravitation.\\
Furthermore, it is of purely quantum origin, i.e. it vanishes for the limit $\hbar\longrightarrow 0$.\\
If one of the radii tends to infinity, we obtain a cylindrical
surface,  particularly, one confined to a quantum wire having the
shape of a plane curve as (seen in Fig.\ref{fig1}). The geometric
potential (surface potential $C$), for such system reads
as\cite{net1}
\begin{equation}
U(s)=-\frac{\hbar ^2}{8 m R^2(s)}.
\label{eq7}
\end{equation}
The Hamiltonian of Eq.(\ref{eq5}) dose not contain $y$ and
the $p_{y}$ is a constant of motion with the value of $\hbar k$.
The wave function $Y(s,y)$ in $y$ direction as planar waves
$e^ {iky}$  and in $s$ direction is as function of $s$, therefore,
we can assume
\begin{equation}
 Y(s,y)=e^ {iky}p(s).
\label{eq8}
\end{equation}
Substituting  Eq.(\ref{eq8}) in  Eq.(\ref{eq5}) we can
obtain the  Hamiltonian of electron gas in two dimensional as
\cite{net25}
\begin{equation}
H=\frac{-\hbar^2}{2m}\frac{\partial^2}{\partial{s}^2}+\frac{1}{2}m\omega_{c}^2 \Big(x(s)-\frac{\hbar k}{m\omega_{c}} \Big)^2-\frac{\hbar^2}{8 mR^2(s)}.
\label{eq9}
\end{equation}
Note that for $f'(x)\ll1$, we can expand Eq.(\ref{eq1}) and by using one order approximation, we have
$$s(x)\approx x(s).$$
Let us define  the coordinates $x'(s)$ and $s'(x)$ as
\begin{equation}
x'(s)=x(s)+\frac{\hbar k}{m\omega_{c}},  \;\;\;\;\;
s'(x)=s(x)+\frac{\hbar k}{m\omega_{c}},
\label{eq10}
\end{equation}
which allow us to rewrite the Eq.(\ref{eq9}) in two parts as
\begin{equation}
H= H_{os}+\frac{1}{2}m\omega_{c}^2 \Big[x'^{2}(s)-s'^{2}(x)\Big]-\frac{\hbar ^2}{8 m R^2(s)},
\label{eq11}
\end{equation}
where $H_{os}$ is Hamiltonian of a simple harmonic oscillator.Note that since the distance parameter on cylinder surface $s(x)$ of Fig.\ref{fig1} is small, it is convenient to consider second and third terms of Eq.(\ref{eq11}),  as a perturbation Hamiltonian.
\section{The energy spectrum}
\begin{figure}[H]
\centerline{\includegraphics[scale=1]{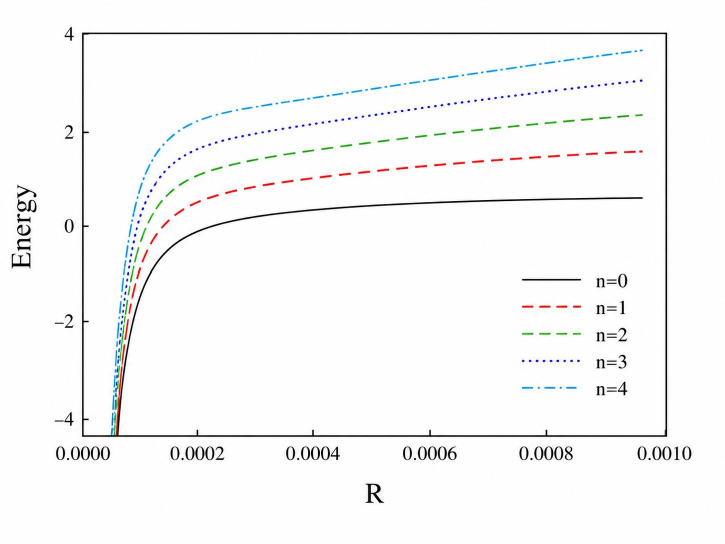}}
\caption{Plot of the energies versus the radius $R$ with different Landau indices $n$, under the effect constrained potential in a curved sample of Eq.(\ref{eq20}); here, $e=1.602 \times 10^{-19} C, m=9.109 \times 10^{-31} kg, c=3 \times 10^{8} m/s,\hbar=1.055 \times 10^{-34} J.s, k=1.75 \times 10^{11} C.T/kg, B=1T.$ In  the  limit $R\rightarrow \infty$ the energies  spectrum have asymptotic behavior.}
\label{fig2}
\end{figure}

In this section, the time independent perturbation theory is used and then  the energy spectrum is calculated, thus we have,
\begin{equation}
H'=\frac{1}{2}m\omega_{c}^2 \Big[x'^2(s)-s'^2(x)\Big] -\frac{\hbar^2}{8 m R^2}.
\label{eq12}
\end{equation}
The first order energy shift is obtained by
\begin{equation}
E_{n}^{1}=\frac{1}{2}m\omega_{c}^2\Big[\langle n|x'^2(s)|n\rangle -\langle n |s'^2(x)|n\rangle  \Big]- \frac{\hbar ^2}{8m R^2}\langle n|n\rangle.
\label{eq13}
\end{equation}
In order to calculate the first and second terms of the Eq.(\ref{eq13}),
 let us define in Fig.\ref {fig1}
\begin{equation}
x'(s)=R \sin(\frac{s(x)}{R})+\frac{ \hbar k}{m\omega_{c}}.
\label{eq14}
\end{equation}
Substituting Eq.(\ref{eq14}) into Eq.(\ref{eq13}) we can find
\begin{eqnarray}\label{eq15}
&&\langle n|x'^2(s)|n\rangle  = R^{2}\Big[\langle n|(-\frac{1}{4}\exp(\frac{2\imath \hat{s}}{R})+\exp(-\frac{2\imath \hat{s}}{R})-2) |n\rangle\Big]\nonumber\\
&&+\frac{\hbar^{2}k^{2}}{m^{2}\omega^{2}_{c}}+\frac{2\hbar kR}{m\omega_{c}}\Big[\langle n|\frac{1}{2\imath}(\exp(\frac{\imath \hat{s}}{R})-\exp(-\frac{\imath \hat{s}}{R})) |n\rangle\Big]\nonumber\\
&&\langle n |s'^2(x)|n\rangle=\frac{\hbar}{2m\omega_{c}}(2n+1).
\end{eqnarray}
It is convenient to define two non-Hermitian operators as
\begin{equation}
\hat{a}=\sqrt{\frac{m\omega}{2\hbar}}(\hat{s}+\frac{i\hat{p}}{m\omega}),\;\;\;\;\;\;\;\;
\hat{a}^{\dag}=\sqrt{\frac{m\omega}{2\hbar}}(\hat{s}-\frac{i\hat{p}}{m\omega}),
\label{eq16}
\end{equation}
where $p=i\hbar \partial/\partial{s}$.\\
 Using the Eqs.
(\ref{eq15}) and (\ref{eq16}), the Eq.(\ref{eq13}) reduce to
\begin{eqnarray}\label{eq17}
&&E_{n}^{1}=\frac{1}{2}m\omega_{c}^2\Big[\frac{R^2}{2}-\frac{R^2}{4}[\langle n|e^{\frac{2i}{R}\sqrt{\frac{\hbar}{2m\omega_{c}}}
(\hat{a}+\hat{a^{\dag}})}|n\rangle e^{\frac{-2i\hbar k}{m\omega_{c} R}}+\langle n| e^{\frac{-2i}{R}\sqrt{\frac{\hbar}
{2m\omega_{c}}}(\hat{a}+\hat{a^{\dag}})} |n\rangle\nonumber\\ && e^{\frac{2i\hbar k}{m \omega_{c} R}}]
+\frac{\hbar ^2 k^2}{m^2 \omega_{c}^2}+\frac{\hbar k R}{ i m \omega_{c}} [\langle n|e^{\frac{i}{R}\sqrt{\frac{\hbar}{2m\omega_{c}}}
(\hat{a}+\hat{a^{\dag}})}|n\rangle e^{\frac{-i\hbar k}{m\omega_{c} R}}-\langle n|e^{\frac{-i}{R}\sqrt{\frac{\hbar}{2m\omega_{c}}}
(\hat{a}+\hat{a^{\dag}})}|n\rangle e^{\frac{i\hbar k}{m\omega_{c} R}}]\nonumber\\ && -\frac{\hbar}{2m\omega_{c}}(2n+1)\Big]-\frac{\hbar ^2}{8m R^2}
 \langle n|n\rangle.
\end{eqnarray}
Let us  define the dimensionless parameters $\alpha$  and
$\beta$ as
\begin{equation}
\alpha=\frac{2}{R}\sqrt{\frac{\hbar}{2 m \omega_{c}}},\;\;\;\;\;\;
\beta=\frac{2\hbar k}{m \omega_{c} R}.
\label{eq18}
\end{equation}
Using the formula
\begin{eqnarray}\label{eq19}
&&\langle n|(\hat{a^{\dagger}})^{\ell}\hat{a}^{m}|n\rangle=\frac{n!}{(n-l)!}\delta_{\ell,m},\nonumber\\
&&e^{\hat{A}+\hat{B}}=e^{\hat{A}}e^{\hat{B}}e^{-\frac{1}{2}[\hat{A},\hat{B}]},\nonumber\\
&&e^{x}=\sum_{0}^{\infty}\frac{x^{n}}{n!},
\end{eqnarray}
we find the energy spectrum for the first order of $n$ of a curved
(2DEG) under a homogenous field in cylindrical geometry as
\begin{eqnarray}\label{eq20}
&&E_{n}=E_{n}^{0}+E_{n}^{1}=(n+\frac{1}{2})\hbar
\omega_{c}+\frac{1}{4}m\omega_{c}^2
R^2\Big[1-e^{-\frac{1}{2}\alpha^2}\cos(\beta)\sum_{m=0}^{n}\frac{(i\alpha)^{2m}}{(m!)^2}\frac{n!}{(n-m)!}\nonumber\\
&&-\frac{4\hbar k }{m\omega_{c}
R}e^{-\frac{1}{4}\alpha^2}\sin(\frac{\beta}{2})\sum_{m=0}^{n}\frac{(i\alpha
/2)^{2m}}{(m!)^2}\frac{n!}{(n-m)!}\Big]\nonumber
+\frac{\hbar^2k^2}{2m}-\frac{1}{4}\hbar\omega_{c}(2n+1)\nonumber\\
&&-\frac{\hbar^2}{8mR^2}.
\end{eqnarray}
 The dependence of 
Eq.(\ref{eq20}) on curvature radius R for a cylindrical surface is
remarkable and has an important consequence. In order to
illustrate the  dynamical properties of Eq.(\ref{eq20}),the energy spectrum as a function of curvature
radius R in the Fig.\ref{fig2} has been plotted for the states $ n=0,1,2,\ldots$.
For small radii R, the drop energy spectrum is observed and for large
radii $R$ ($R\longrightarrow \infty$) the energy spectrum is
independent of $R$ and the energy levels have an asymptotic behaviour as the radius $R$ increases.\\
Note that Eq.(\ref{eq20}) for $R\longrightarrow \infty$
deduces to
\begin{equation}
E_{n}=(n+\frac{1}{2})\hbar\omega_{c}.
\label{eq21}
\end{equation}
This equation has a common expression (in landau gauge) for a flat
2DEG in a perpendicular magnetic field \cite{net13,net14}. In
order to evaluate the results obtained in  Eq.(\ref{eq20}), we have
compared Fig.\ref{fig2} with plot3 and plot5 which were obtained numerically
in \cite{net26}. These figures show that the energy levels versus
the radius $R$ of the cylinder have an asymptotic behaviour as 
$R$ increases. In the limit $R\rightarrow \infty$, they become the
Landau levels for an
electron in a flat space \cite{net26}.\\
Due to technological progress, the physics of curved
two-dimensional quantum system with cylindrical symmetry
 is important  both theoretically and experimentally \cite{net29}.
 Therefore, with  study of  energy spectrum
of  these systems, we canninvestigate the role of curvature in
electronics as magnetotransport \cite{net24} and quantum
electromechanical circuits \cite{net9}. Using the Eq.
(\ref{eq20}), we will study in future works of  the magnetic properties
of curved two-dimensional electron gas as the chemical potential,
ground-state energy and magnetic susceptibility.
\section{Curved 2DEGS in high magnetic fields}
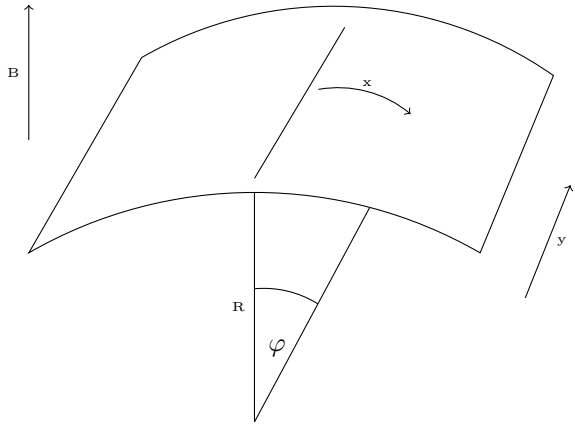
\begin{figure} \label{fig3}
\centering
\begin{tikzpicture}[scale=3]
\draw (1,-.75) -- (1,.27)node[pos=0.5,left]{\tiny R};
\draw (1,.33) -- (1.4,1);
\draw (1,-.75) -- (1.51,.2);
\draw[->] (2.2,-.2) -- (2.4,.3)node [pos=0.5,right]{\tiny y};
\draw[->] (0,.5) -- (0,1.1)node[pos=0.5,left]{\tiny B};
\draw [<-](20:1.8) arc (50:100:.5);
\fill (1.5,.7) node[above] {\tiny x};
\draw (-10:1.3) arc (58:95:.45);
\fill (1.1,-.5) node[above] {$\varphi$};
\draw (0:2cm) -- (0:2cm)
arc (60:120:2cm) -- (60:1cm)
arc (120:55:1.7cm) -- cycle;
\end{tikzpicture}
\caption{A two-dimensional electron gas in a magnetic field } 
\end{figure}

Curved 2DEGs in magnetic fields without surface
potential(geometric potentials)in high magnetic fields are studied
in \cite{net15}. In this section, we apply this approach and
consider  influence of curvature (term relevant to the surface
potential)directly in the equation of motion (the
Schrodinger equation). The energy spectrum will be found
for a two dimensional electron gas under the a magnetic field. For
this purpose, we assume that two-dimensional electron gas in
component  plane  of the magnetic field does not influence the
energetic structure of the system  strictly. The geometry of a
curved 2DEG in a homogeneous magnetic field is shown in Fig.\ref{fig3}.The
Schrodinger Eq.(\ref{eq2})  for cylinder 2DEG is given by
\begin{equation}
 \frac{1}{2m}\Big[({\mathbf{P}}-\frac{e}{c}{\mathbf{A}})^2-\frac{\hbar^{2}}{8mR^{2}}\Big]\psi(x,y)=E\psi(x,y).
\label{eq22}
\end{equation}

\begin{figure}[H]
\includegraphics[width=1.5\textwidth]{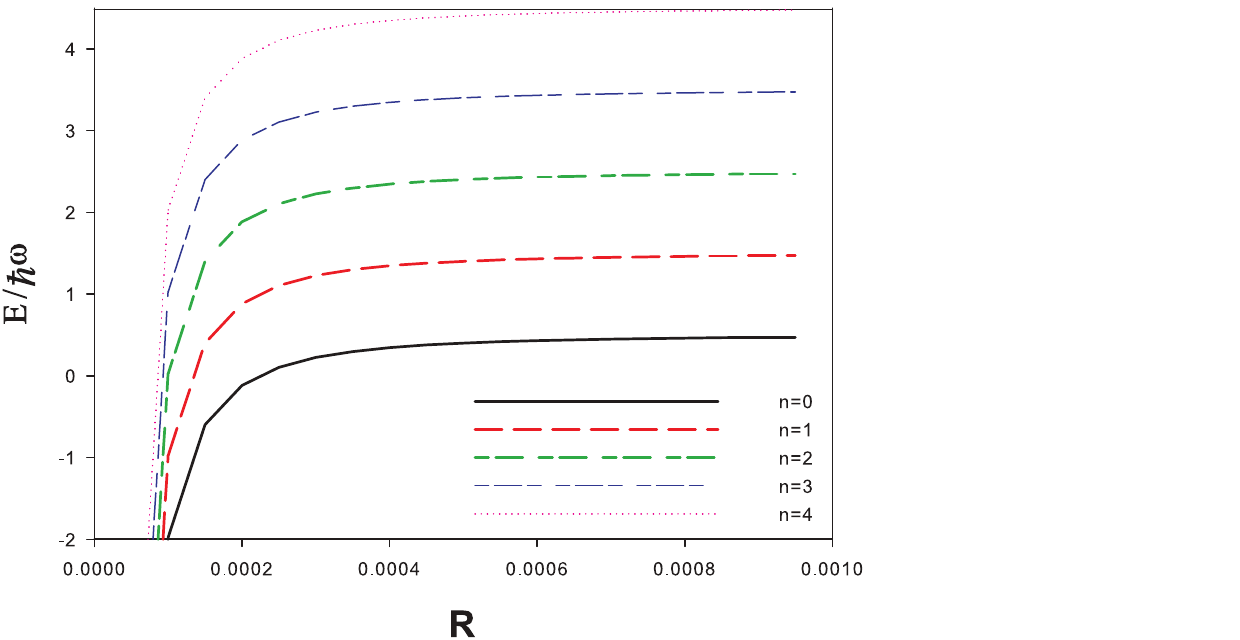}
\caption{Plot of the energies versus the radius $R$ with different Landau indices $n$, in a  curved sample of Eq.(\ref{eq28}); here, $e=1.602 \times 10^{-19} C, m=9.109 \times 10^{-31} kg, c=3 \times 10^{8} m/s,\hbar=1.055 \times 10^{-34} J.s, k=1.75 \times 10^{11} C.T/kg, B=1T.$ The  energy levels have an  asymptotic behavior as the radius $R$ increases. In the limit $R\rightarrow \infty$ they become the landau levels which in the planar case are dispersionless.}
\label{fig4}
\end{figure}

According  to the  Fig.\ref{fig3},
we consider a magnetic field perpendicular to the cylinder as
$\vec{B} = (0,0,Bcos\varphi)$, with an intensity $B$, and a
convenient choice for the vector potential in asymmetric gauge  is
$\vec{A}= (0,BRcos\varphi,0)$.
 Therefore,  substituting  vector potential in the Eq.(\ref{eq22}) and after some mathematical calculations, we find
\begin{equation}
\frac{-\hbar^2}{2m} \frac{\partial^2{\psi(x)}}{\partial{x}^2}+\frac{1}{2} m\omega_{c}^2(R\sin\frac{x}{R}-\frac{\hbar k_{y}}{eB})^2\psi(x) -\frac{\hbar^2}{8mR^2}\psi(x) =E \psi(x),
\label{eq23}
\end{equation}
where $\omega_{c}=e B/mc$, is the cyclotron frequency and $k_{y}$ is the wave vector of motion
in y-direction.
Note that the Eq. (\ref{eq23}) for
$R\longrightarrow \infty$ reduces to the flat 2DEG in a
perpendicular magnetic field as
\begin{equation}
\frac{-\hbar^2}{2m} \frac{\partial^2{\psi(x)}}{\partial{x}^2}+\frac{1}{2} m\omega_{c}^2(x-x_{0})^2\psi(x) =E\psi(x),
\label{eq24}
\end{equation}
where $ x_{0}=\hbar k_{y}/e B$ is the apex of parabola that is created by magnetic field.
Eq.(\ref{eq23}) can be simplified by presentating local coordinates around  the minimum $x_{m}=R \arcsin(x_{0}/R)$ of the term in parenthesis
\begin{equation}
\sin\frac{x_{m}+x'}{R}-\frac{x_{0}}{R}=\sin\frac{x_{m}}{R}\underbrace{\cos\frac{x'}{R}}_{\approx 1}+\cos\frac{x_{m}}{R}\underbrace{\sin\frac{x'}{R}}_{\approx x'/R}-\frac{x_{0}}{R}\approx\frac{x'}{R}\cos\frac{x_{m}}{R}.
\label{eq25}
\end{equation}
Here we assume that the local coordinate $x' \ll R$. Since the radius of curvature $R$ is of the order $10^{-4}m$ in Fig.\ref{fig4}, whereas the local coordinate is relevant on the scale of the magnetic length $(\approx 25\;nm\; at\;1\; T),$ this assumption is well satisfied \cite{net6, net15, net16, net17}.\\
Substituting Eq.(\ref{eq25}) into the Eq.(\ref{eq23}), we have
\begin{equation}
\frac{-\hbar^2}{2m}\frac{\partial^2{\psi(x)}}{\partial{x^2}} +\frac{1}{2} m(\omega_{c}\cos\frac{x_{m}}{R})^2(x-x_{m})^2\psi(x) -\frac{\hbar^2}{8mR^2}\psi(x) =E\psi(x),
\label{eq26}
\end{equation}
which is equivalent to Eq.(\ref{eq24}) only with difference
that the effective magnetic field is $B_{eff}=B \cos(x_{m}/R)$,
we find energy spectrum of a curved 2DEG in cylindrical
geometry  as
\begin{eqnarray}
&&E_{n}=(n+\frac{1}{2})\hbar\omega_{c}\cos\frac{x_{m}}{R}-\frac{\hbar^2}{8mR^2}=(n+\frac{1}{2})\hbar\omega_{c}\cos(\arcsin\frac{x_{0}}{R})-\frac{\hbar^2}{8mR^2} \nonumber\\
&&E_{n}=(n+\frac{1}{2})\hbar\omega_{c}\sqrt{1-\frac{x_{0}^2}{R^2}}-\frac{\hbar^2}{8mR^2}.
\label{eq27}
\end{eqnarray}
Eq. (\ref{eq27}) shows  that the landau levels  which  are dispersionless in
the planar case, now have a semielliptical
dispersion as a function of $R$ and $B$. The Eq.(\ref{eq27}) is
valid for the energy spectrum of a single-particle state of a
two-dimensional electron gas confined to the surface of a cylinder
immersed in a magnetic field.  In Fig.\ref{fig4},
energy spectrum is shown curvature plots for the states $n =0, 1,\ldots$. For small
radii $R$, the low energy spectrum is observed and for large radii $R$
($R\longrightarrow \infty$), it is independent of, $R$ i.e. the
flat 2DEG. In order to evaluate the validity of Eq. (\ref{eq20}), we
 are comparing Figs.\ref{fig2} and \ref{fig4}. It can be
concluded  that the dependence to curvature $R$ for Fig.
\ref{fig2} is more. This dependence can also be
seen for some values of the dimensionless parameters $\alpha$ and
$\beta$ in Eq. (\ref{eq20}). The energy spectrums in Eqs. (\ref{eq20}) and (\ref{eq27}) are identical in the limit state $R\longrightarrow \infty$. In this state, these equations show the Landau levels for a flat 2DEG in a perpendicular magnetic field.\\
In this work, we investigate the bound states of a quantum
particle on the curved surface with cylindrical symmetry in the
context of the Schrodinger theory considering  da
Costa's approach. In another approach, the study of bound state is
based on  Klein-Gordon type equation on surfaces, without 
constraining potential \cite{net30,net31}. In this method, the
effective potential is given by
\begin{equation}
V_{D}=\frac{\hbar ^2}{2m}K.
\label{eq28}
\end{equation}
 Where K is Gaussian curvature
of the surface. Since K = 0 on the cylinder surface 
 \cite{net32}, Klein-Gordon type equation is independent of
$R$.
\section{conclusion}
In this work, the effect of the potential that constrains the
particle to the surface has been considered. For curved samples of
nanostructures, the effect of the curvature
 on the energy spectrum of a two-dimensional electron gas has been
obtained perturbatively. The quantum particles transport in curved
waveguide is described by a Hamiltonian consisting of the kinetic
energy operator and a resulting potential energy, which is of pure
geometric origin. Thus it is
worthwhile studying the influence of geometric potentials
(Curvature) on propagating particles in curved low-dimensional
electron systems. \normalsize \small { \baselineskip=.5cm
\bibliographystyle{./styles/plain-fa}
\providecommand{\noopsort}[1]{}
~\\

\end{document}